\documentclass[3p,times]{elsarticle}

\usepackage{ecrc}
\usepackage[cmex10]{amsmath}
\usepackage{array}

\volume{00}

\firstpage{1}

\journalname{Computational Materials Science}

\runauth{M.E. Semenov et al.}

\CopyrightLine{2011}{Published by Elsevier Ltd.}

\usepackage{amssymb}

 \biboptions{sort&compress}

\usepackage[figuresright]{rotating}

\begin{document}

\begin{frontmatter}



\dochead{}

\title{Computational investigation of plastic deformation \\
in face-centered cubic materials}



\author[sme]{Mikhail Semenov\corref{cor1}
}
\ead{sme@tpu.ru}
\author[ksn]{Svetlana Kolupaeva
}
\cortext[cor1]{Corresponding author} 
\address[sme]{Department of High Mathematics and Mathematical Physics, Tomsk Polytechnic University, 30, Lenin ave., Tomsk, 634050, Russia}
\address[ksn]{Department of Applied Mathematics, Tomsk State University of Architecture and Building, 2, Solyanaya sq., Tomsk, 634003, Russia}


\begin{abstract}
A mathematical model of plastic deformation in face-centered cubic
(FCC) materials based on a balance model taking into account
fundamental properties of deformation defects of a crystal lattice
was developed. This model is based on a system of ordinary
differential equations (ODE) accounting for various mechanisms of
generation and annihilation of deformation defects for different
external conditions. In-house developed software, SPFCC (Slip
Plasticity of Face-Centered Cubic), was employed to solve the
system of ordinary differential equations. The implemented code
solves efficiently the stiff ODE system and provides a
user-friendly interface for investigation of various features of
plastic deformation in FCC materials.

Simulation of plastic deformation in the FCC metals was performed
for the case of constant strain rate. The modelling results were
validated by comparing experimental data and simulation results
(stress-strain curves) and good agreement was obtained.

\end{abstract}

\begin{keyword}
Deformation defects \sep Flow stress \sep Mathematical model \sep
Plastic deformation \sep Software \sep Stiff ODE


\end{keyword}

\end{frontmatter}


\section{Introduction}

The role and possibilities of mathematical modelling and computer
simulation in the analysis of plastic deformation are of great
importance. The mathematical modelling allows to reveal the role
of various factors influencing processes of plastic deformation
such as mechanisms responsible for strain hardening and evolution
of deformation defects of crystalline materials
\cite{Kuhlmann1962, Preston2003, Kocks03, Kolupaeva10, Zhan2011}.
Furthermore, the mathematical modelling is increasingly utilised
for planning of experimental studies.

The mathematical modelling of plastic deformation in crystals is
based on mathematical relationship taking into account fundamental
properties of deformation defects of crystal lattice and
represents a fundamental part of crystal plasticity models
\cite{Kolupaeva10, Lagneborg1972, Essmann1979, Mecking1981,
Estrin1998, Popov2000}. The models usually consist of kinetic
equation, which establishes a relationship between the current
microstructure and the material's mechanical response. This
relationship,is written in general form as ordinary differential
equations (ODE). Various models reported in literature
\cite{Zhan2011, Lagneborg1972, Mecking1981, Zerilli1987,
Voyiadjis2008, Chinh2010} differ from each other in way the set of
deformation defects and the mechanisms of their generation and
annihilation are considered.

Typically mathematical models of plastic deformation include the
equations of deformation defects accounting for elementary
processes and mechanisms of plasticity \cite{Kolupaeva10,
Lagneborg1972, Essmann1979, Popov2000a}. Specifically, the choice
of deformation defects, defining the regularities of plastic
deformation and the mechanisms of their generation and
annihilation and mutual transformation, determines the adequacy
and possibilities of a mathematical model.

Crystallographic shear zone is selected as a basic structural
element with which regard mechanisms of slip plasticity are
considered in mathematical model of plastic deformation by slip
\cite{Kolupaeva10, Popov2000, Popov2000a, Orlov1966}. The
description of the mechanisms of formation of shear zones
formation is based on the fundamental physical and topological
properties of defects responsible for plastic mass transfer. The
fact that all parameters of the model either have physical or
geometric interpretation or can be defined from physical
considerations, or their values can be limited to a certain
interval of may be specified, is a characteristic feature of the
model and represent a significant advantage over phenomenological
models.

Plastic deformation of slip primarily produces dislocations of
various types and point defects. The processes of initiations of
defects during formation of crystallographic shear zones are the
mechanisms defining their generation.

Mathematical model of plastic deformation in FCC metals
\cite{Kolupaeva2005, Semenov2010} consists of the balance
equations of shear-producing dislocations (their density is
denoted as $\rho_m$), dislocations in vacancy ($\rho_d^v$) and
interstitial ($\rho_d^i$) dipole configurations, interstitial
atoms (concentration $c_i$), monovacancies ($c_{1v}$) and
divacancies ($c_{2v}$), the equation defining strain rate, and the
equation describing external effects to the deformed material. The
specific ingredients of the model of deformation defects
generation and annihilation are derived from the assumption that:
\begin{enumerate}
\item The defects generation processes occur during the formation
of crystallographic shear zones and are related to the dynamic
nature of their formation. \item Annihilation processes have
predominantly diffusion nature and occur in defect medium created
by a set of defects generated by the slip in a large number of
shear zones. \item Deformation defect continuum is homogeneous
continuum and contains the same number of defects, that the entire
crystallographic shear zones together. \end{enumerate}

The mechanisms of deformation defects generation during the
formation of crystallographic shear zone in monocrystals of FCC
metals are reviewed in section~2, and the mechanisms of defects
annihilation are presented in section~3. Section~4 summarises the
system of differential equations of the model of plastic
deformation in FCC metals. Section~5 is devoted to the description
of SPFCC software, aimed to automatize the investigation of
regularities of plastic deformation by slip with the use of a
mathematical model. Section~6 gives an illustrative example
showing possibilities of SPFCC software.

\section{The intensity of deformation defects generation in the formation of crystallographic shear zone}

\textbf{The intensity of shear-producing dislocations generation.}
The intensity of the generation of shear-producing dislocations
during the formation of crystallographic shear zone was obtained
in \cite{Popov1982} based on the ideas of Indenbom~V.L.,
Orlov~A.K., and Kuhlmann-Wilsdorf~D. \cite{Kuhlmann1962,
Orlov1966, Indenbom1966}. If we assume that the process of
deformation occurs in identical shear zones, the intensity of the
generation of shear-producing dislocations with deformation can be
described as follows \cite{Popov1982}:
\begin{equation}\label{eq1}
G_m(\rho)=F/(bD_r),       
\end{equation}
where $D_r=(B_r \tau)/(G b\rho_m)$ is an average diameter of the
shear zone, $b$ is the magnitude of the Burgers vector, parameter
$F$ is determined by the geometry of the dislocation loops and
their distribution in the shear zone \cite{Popov1982}, $B_r$ is
the parameter determined by the probability of formation of
extended dislocation barriers, $\tau$  is the applied stress, $G$
is the shear modulus. The description of the parameters and
variables of the mathematical model is summirised in
Tabl\ref{table1} (see Appendix~A).

Dislocations in dipole configurations are generated during the
formation of crystallographic shear zone. The main cause for the
occurrence of these dislocations is the presence of immobile
dislocation jogs. When moving, the jogs leave a chain of point
defects in case the height of the jogs is equal to one interatomic
distance. In cases where the height of the jogs is more than one
interatomic distance a dipole is formed. The dislocation loops
emitted by the source broaden rapidly, due to decrease in nonlocal
curvature. This leads to the generation of point defects with
increasing intensity, and orientation close to screw, and their
entrapment after short period of time.
\begin{equation}\label{eq2}
G_{di}(\rho) = G_{dv}(\rho) = (6\gamma_d l_d(\rho)b)^{-1}, 
\end{equation}
where $\gamma_d$ is the size parameter of the particles,
$l_d(\rho)$ is the length of the screw dislocation run.

\textbf{The intensity of generation of point defects.}
Non-conservative movement of dislocations with jog leads to the
generation of such point deformation defects as interstitial
atoms, vacancies and divacancies. The intensity of generation of
point defects can be calculated as
\begin{equation}\label{eq2a}
G_k(\rho)= 2q \tau_{dyn}/G,
\end{equation}
where $k=\{i, 1v, 2v\}$, $\tau_{dyn}$ is excessive stress (the
stress excess over the static resistance to dislocation motion),
and parameter $q$ characterizing local concentration of stress
during shear zone formation.

If the distribution of forest dislocations is assumed random, and
the intensity of vacancies and interstitial atoms generation is
considered identical, the relation for the rate of the vacancies
and the interstitial atoms generation is as follows:
\begin{equation}\label{ed3}
G_i(\rho) = G_{1v}(\rho) = q \tau_{dyn}/G.                                  
\end{equation}

Taking into account the fact that the activation energy of
divacancies migration is lower than the activation energy of
monovacancies migration, and that divacancies migrate from
vacancies chain behind threshold on dislocations to crystal volume
presumably faster, the following relations for the intensity of
mono- and divacancies generation during the process of deformation
were derived:
\begin{equation}\label{eq4}
G_{1v}(\rho) = q \tau_{dyn}/(6G),      
\end{equation}
\begin{equation}\label{eq5}
G_{2v}(\rho) = 5q \tau_{dyn}/(6G).       
\end{equation}

\section{Mathematical modelling of deformation defects annihilation processes}

\textbf{Dislocations annihilation.} A number of different
mechanisms of dislocations annihilation have been presented in the
literature \cite{Kolupaeva10, Lagneborg1972, Essmann1979,
Popov2000, Kolupaeva2005, Popov1982}. The main mechanisms of
dislocations annihilation are: a) climb of non-screw dislocations
as a result of precipitation of point defects on extra planes of
non-screw dislocations, and b) annihilation of screw dislocations
during their cross-slip.

The following mechanisms are considered for annihilation of dipole
dislocation configurations:
\begin{enumerate}[1.]
\item Decrease in the shoulder of vacancy dislocation dipoles
until their annihilation with precipitation of interstitial atoms
onto them. \item Decrease in the shoulder of interstitial
dislocation dipoles until their annihilation with precipitation of
monovacancies and divacancies onto them. \item Increase of the
shoulder of vacancy dislocation dipoles until the loss of their
stability with precipitation of monovacancies and divacancies onto
them. \item Increase of the shoulder of interstitial dislocation
dipoles until the loss of their stability with precipitation of
interstitial atoms onto them.
\end{enumerate}

\textbf{Shear-producing dislocations annihilation by climb.}
Annihilation processes of diffusive nature in a dislocation
subsystem include both climb of non-screw dislocations, and their
counter-movement in the slip planes. Dislocations annihilation by
crystallographic slip can occur only in pairs of dislocation
segments, which are located closely enough to one another and have
opposite Burgers vectors. The distance between such dislocation
segments is defined by critical radius of capture:
\[
r_a=\frac{Gb}{4\pi\tau_f}\frac{2-\nu}{1-\nu},
\]
where $\tau_f$ is friction stress, $\nu$ is Poisson's ratio.

At high density of dislocations when the distance between
dislocations is less than $r_a$, it can be assumed that the
distance between dislocation segments able to annihilate is equal
to $\rho_m^{-1/2}$. Assuming that dipoles are distributed
uniformly along the length of the shoulder and the distance
between the dipoles in interval $[0, \rho_m^{-1/2}]$, then the
expression for the rate of the shear-forming dislocations
annihilation as a result of precipitation of interstitial atoms
onto them can be written as:
\begin{equation}\label{eq6}
A_m(\rho, c_k) = 2b^{-1}\rho_m min(r_a,  \rho_m^{-1/2}) \omega_m^i / A_i(\rho, c_k), 
\end{equation}
where $A_i(\rho, c_k)=Q_kc_k/\dot{a}$ is the rate of point defects
annihilation, $k=\{i, 1v, 2v\}$, $Q_k$ is the diffusion
coefficient, $\dot{a}$ is the strain rate,  $\omega_m^i$ is the
fraction of interstitial atoms which left the glide plane on which
they are stored.

\textbf{Screw dislocations annihilation by cross-slip.} We assume
that the cross-slip of dislocations occurs only at temperatures
above a certain temperature $T_{cs}$ and is athermal. In this
case, almost all screw dislocations annihilate if they are stable
under current applied stress $\tau$ and are trapped in dipole
configurations in which the distance between dislocations of the
opposite sign does not exceed the distance $r_a=Gb/(2\pi\tau_f)$.
According to this approach the annihilation of screw dislocations
by cross-slip can be represented by the relation:
\begin{equation}\label{eq7}
A_m^{cs}(\rho_m) = FG \rho_m \omega_s P_{as}/(B\tau),                                  
\end{equation}
where $P_{as}$ is the probability of annihilation of screw
dislocations, $\omega_s$ is a fraction of screw dislocations.

\textbf{Dislocations annihilation in dipole configurations.} The
main mechanism of dislocation annihilation in dipole configuration
is the trapping of interstitial atoms on the vacancy dipoles,
which leads to a decrease of the shoulder of the vacancy dipoles,
and precipitation of interstitial atoms on interstitial dipoles.

\textbf{Dislocations annihilation in vacancy dipole
configurations.} Suppose that the maximum shoulder of the dipole,
after which it loses its stability, is equal to the critical
radius of the capture $r_a$. In this case the rate of the
dislocations annihilation in vacancy dipole configurations can be
determined as follows:
\begin{equation}\label{eq8}
A_{dv}(\rho, c_i) = 2\omega_i^{dv}A_i(\rho, c_i)/(b r_a). 
\end{equation}
Here $\omega_i^{dv}$ is the fraction of interstitial atoms, which
left to vacancy dipoles.

The density of dislocations forming vacancy dipoles during
precipitation of vacancies onto them does not change significantly
until the length of dipole shoulder becomes higher than $r_a$. In
this instance, dipoles loose stability and their further behaviour
become analogous to that of the shear-forming dislocations. The
rate of change of the density of dislocations in vacancy dipole
configurations with deformation caused by precipitation of
vacancies can be described as follows:
\begin{equation}\label{eq9}
R_{dv}(\rho, c_{1v})=2\omega_{1v}^{dv} A_{1v}(\rho, c_{1v})/(b r_a), 
\end{equation}
where $\omega_{1v}^{dv}$ is the fraction of vacancies, which left
to vacancy dipoles.

The rate of change of the dislocation density in vacancy dipoles
configurations connected with precipitation of monovacancies onto
them can be described analogically:

\begin{equation}\label{eq10}
R_{dv}(\rho, c_{2v}) = 2\omega_{2v}^{dv} A_{2v}(\rho, c_{2v})/(b
r_a), 
\end{equation}
where $\omega_{2v}^{dv}$ is the fraction of divacancies, which
left to vacancy dipoles.

\textbf{Dislocations annihilation in interstitial dipole
configurations.} Taking into account the fact that precipitation
of vacancies onto interstitial dislocation dipoles leads to the
decrease of the shoulder of the interstitial dislocation dipoles,
whereas precipitation of interstitial atoms leads to the increase
in their shoulder, similar to the intensity of the annihilation of
vacancy dipoles considered above Eq.~(\ref{eq8}), we can write
down the following relations for the intensity of the annihilation
of interstitial dipole configurations in case of precipitation of
mono- and divacancies onto them respectively:
\begin{equation}\label{eq11}
A_{di}(\rho, c_{1v}) = 2\omega_{1v}^{di}A_{1v}(\rho, c_{1v})/(b r_a), 
\end{equation}
\begin{equation}\label{eq12}
A_{di}(\rho, c_{2v}) = 2\omega_{2v}^{di}A_{2v}(\rho, c_{2v})/(b
r_a), 
\end{equation}
where $\omega_{1v}^{di}$, $\omega_{2v}^{di}$ are the fractions of
mono- and divacancies respectively, which left to interstitial
dipoles.

For the intensity of change of the dislocations density in
interstitial dipole configurations, which is connected with
precipitation of interstitial atoms onto them, the following
relation is used:
\begin{equation}\label{eq13}
R_{di}(\rho, c_i) = 2\omega_i^{di}A_i(\rho, c_i)/(b r_a), 
\end{equation}
where $\omega_i^{di}$ is a fraction of interstitial atoms, which
left to interstitial dipoles.

\textbf{Annihilation of deformation point defects.} During the
process of plastic deformation, concentration of point defects is
reduced as a result of their spending for the sinks. Dislocations,
crystal grain boundaries, and crystal surface can act as the sinks
for point defects. In this work, the following sinks are
considered in particular models of point deformation defects
annihilation: 1) for interstitial atoms -- non-screw dislocations,
mono-, divacancies; 2) for monovacancies -- non-screw
dislocations, interstitial atoms, monovacancies; 3) for
divacancies -- non-screw dislocations, interstitial atoms. The
rates of change of the deformation defects concentration can be
expressed as follows:
\begin{equation}\label{eq14}
A_k(\rho, c_k)=Q_k c_k /\dot{a}.                                      
\end{equation}

It should be noted that, when two vacancies in their random
movement meet, divacancies are formed, and when a divacancy meets
an interstitial atom, or an interstitial atom meets a divacancy, a
monovacancy is formed. Therefore, relations of deformation point
defects accumulation take into account an additive to
monovacancies and divacancies accumulation:
\begin{equation}\label{eq15}
R_{1v}(c_{2v}, c_i) = (Q_i +Q_{2v})c_ic_{2v}/\dot{a}, 
\end{equation}
\begin{equation}\label{eq16}
R_{2v}(c_{1v}) = Q_{1v}c_{1v}^2/\dot{a}.                               
\end{equation}

The cases of point defects formed as a result of the meeting of
two divacancies or a mono- and a divacancy were not considered in
this work.

\section{The system of differential equations of deformation
defects balance}

The presented models of deformation defects generation and
annihilation described in sections~2--3 provide the basis for the
creation of model of plastic deformation by slip. This model can
be used to analyse the influence of various mechanisms and
processes for FCC metals under various conditions of deformation.

The system of differential equations of the balance of deformation
defects for FCC metals can be summarized as follows
\cite{Kolupaeva10}:
\begin{align}\label{eq17}
      \displaystyle
      \frac{dc_{1v}}{da}
      &= \frac{q \tau_{dyn}}{6G} - \frac{(Q_{1v}c_{1v}- (Q_i+Q_{2v})c_i c_{2v})}{\dot{a}},  \\
      \frac{dc_{2v}}{da}
      &= \frac{5q \tau_{dyn}}{6G} - \frac{(Q_{2v} c_{2v} - Q_{1v}c_{1v}^2)}{\dot{a}}, \nonumber \\
      \frac{dc_i}{da}
      &= \frac{q \tau_{dyn}}{G} - \frac{Q_ic_i}{\dot{a}}, \nonumber \\
      \frac{d\rho^v_d}{da}
      &= \frac{1}{6\gamma_dl_d(\rho)b} - \frac{2(\omega_i^{dv}A_i(\rho, c_i) + \omega_{1v}^{dv}A_{1v}(\rho, c_{1v})+
      \omega_{2v}^{dv}A_{2v}(\rho, c_{2v}))}{b r_a}, \nonumber\\
      \frac{d\rho^i_d}{da}
      &= \frac{1}{6\gamma_d l_d(\rho)b}- \frac{2(\omega_{1v}^{di}A_{1v}(\rho, c_{1v}) +
       \omega_{2v}^{di}A_{2v}(\rho, c_{2v}) + \omega_{id}^{i}A_i(\rho, c_i))}{b r_a}, \nonumber \\
      \frac{d\rho_m}{da}
      &= \frac{F}{bDr} - \frac{FG \rho_m \omega_s P_{as}}{B\tau} - \frac{2\rho_m min(r_a,\rho_m^{-1/2})}{b}\sum_{k=\{i, 1v, 2v\}} \frac{\omega_m^k}{A_k(\rho, c_k)} + \nonumber \\
      &+\frac{2(\omega_{1v}^{dv}A_{1v}(\rho,c_{1v})+\omega_{2v}^{dv}A_{2v}(\rho, c_{2v})+ \rho_i^{di}A_i(\rho, c_i))}{b r_a} \nonumber.
\end{align}

The last equation of the system (\ref{eq17}) takes into account
the relaxation contribution due to the rate of change of the
dislocations density in dipole configurations for the intensity of
shear-forming dislocations accumulation.

The equations establishing the relationship between strain rate
the applied stress as well as the defect condition of the material
are taken from \cite{Kolupaeva10, Popov1982}. In this work the
following equation for the strain rate of plastic deformation is
selected \cite{Kolupaeva10}:
\begin{equation}\label{eq18}
\dot{a}=a_0 \exp\left(-\frac{0,2Gb^3-(\tau-\tau_a)\lambda(\tau,\rho)b^2}{k T \rho^{1/2}}\right), 
\end{equation}
where $\tau_a$ is an athermal component of stress, $\lambda(\tau,
\rho)$ is the function of stress and density of dislocations,
$a_0$ is pre-exponential factor.

The system of equations (\ref{eq17}) should be added by an
equation (or equations), describing the influence causing
deformation.

\section{SPFCC software description}

A computational tool, denoted SPFCC (Slip Plasticity of
Face-Centered Cubic), has been developed. This software can assist
researchers in  application of the mathematical models for the
computational investigation of plastic deformation in FCC
materials. Although, a number of commercial tool addressing this
problem are readily available, a user needs to be familiar with
methods of solving ODE and requires programming language skills in
order to consider a specific problem.

SPFCC software makes it possible to automatically form and
calculate a full or reduced suite of defects of a mathematical
model with little user intervention to set up the input values,
run the calculations, plot stress strain curves, and export the
results \cite{Popov1982}. SPFCC enables direct simulations of
strain hardening and evolution of deformation defects of
crystalline materials.

Presently, the following types of effects can be considered:
deformation with a constant strain rate $\dot{a}=const$, at
constant stress $\tau=const$ and constant load $P=const$ in
conditions of tension $\tau=\tau_0 exp(a/k_s)$ and compression
$\tau=\tau_0 exp(-a/k_s)$, where $k_s$ is Sachs factor.

\begin{figure}[!t]
\centering
\includegraphics[width=6in]{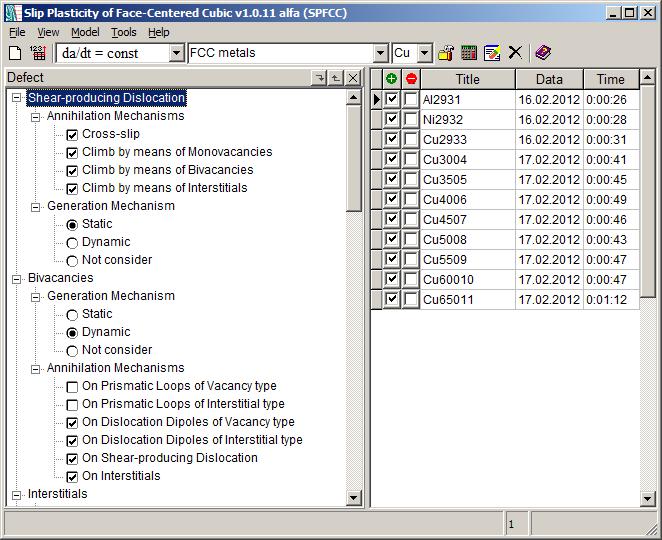}
\caption{Main window of SPFCC software.} \label{fig1}
\end{figure}

Fig.~\ref{fig1} shows main window of the software SPFCC.
Simulation results (stress-strain curves) for different external
conditions are obtained in  less than one minute and posses a
relative accuracy of the solution $10^{-3}$. Fig.~\ref{fig1}
illustrates  the time of simulation results for constant strain
rate at different temperatures. It should be noted that the
computational time for the case of constant stress (creep)
increases and the results are obtained in ten minutes.  All the
simulations were performed on the Supermicro computing cluster
based on Intel Xeon Processor~5570 (code named ``Nehalem'',
http://www.cluster.tpu.ru).

SPFCC software helps to analyse the processes of plastic
deformation in FCC metals and dispersion-strengthened materials
with a metal FCC matrix and noncoherent, undeformable particles of
the second phase \cite{Kolupaeva10}.

The user can either apply the full mathematical model, or create
reduced models. The reduced model can take into account the
deformation defects and mechanisms of their generation and
annihilation specified by the user. At the beginning of a
calculation, the starting suite of defects and initial value of
parameters are selected from the database.

It should be noted that systems of ODE balance of deformation
defects for various deformating effects and materials may differ
in the number of equations and the mechanisms of defects
generation and annihilation because user can to choose or
disregard some mechanisms in the reduced models. In addition, the
equations can be piecewise cross-linked type. This places some
restriction on the choice and realization of a numerical method.

The system of differential equations of the model
(\ref{eq17})--(\ref{eq18}) is numerically stiff \cite{Semenov2010,
Gear1971}. This is mainly due to the fact that processes of
generation and annihilation of deformation defects are
significantly differ in terms of intensity. In addition value of
variables are changing in the interval of integration by orders of
magnitude.

In order to solve a stiff ODE system, the software package SPFCC
employs the explicit linear multistep Adams method to obtain
initial values and variable order multistep Gear method (in the
form of backward differentiation formulas). The Gear method is
stable at any integration stepsize, and therefore  can choose an
integration step guided only by considerations of accuracy, but
not stability \cite{Gear1971}. The original Gear method has been
modified \cite{Semenov2010}, due to some physical restrictions
imposed on the variables of the equations system (for example,
variables can not be negative). The analysis of the effectiveness
of numerical algorithms for the case of plastic deformation of
dispersion-strengthened materials with undeformable particles was
given in publication \cite{Semenov2010}.

The graphical user interface of the SPFCC software allows to
select variables and mechanisms of the generation and annihilation
of dislocations (Fig.~\ref{fig1}). The user can chose values of
the mathematical model parameters, initial conditions, and
calculations accuracy in a dialogue mode. Furthermore, two
material FCC metal or dispersion-strengthened material can be
selected by the user.I n this case, the user will automatically be
offered to choose the values of the parameters of material from
the database software. It should be noted that the database is
compiled using the results of theoretical and experimental studies
available in literature. The user can accept the default values or
set its own parameters values. If a user enters an incorrect
values the SPFCC software will generate an error message and will
prompt user to change value or to set the default values.

The user can perform a series of calculations with the help of
SPFCC software. To run the calculation, the parameter of the
mathematical model (stress or load, temperature, initial density
of dislocations, etc.), needs to be chosen together with the
interval and the step of this parameters.

\begin{figure}[!t]
\centering
\includegraphics[width=6in]{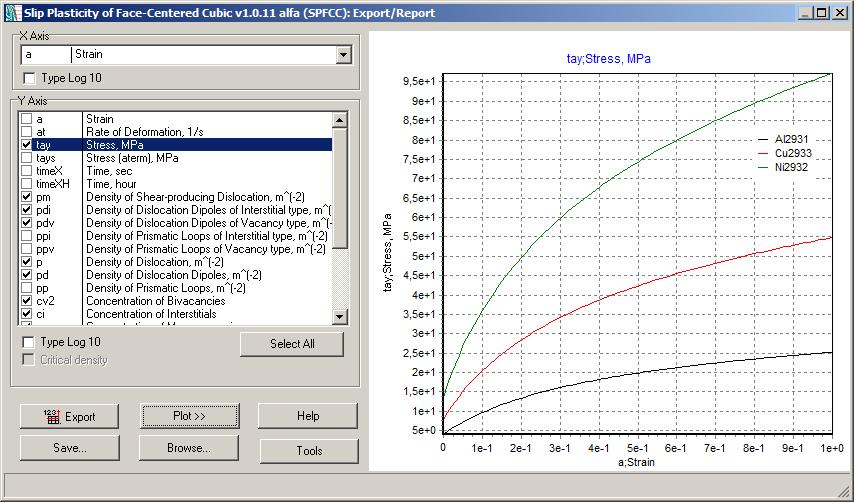}
\caption{Dialogue window ``Export/Report''.} \label{fig2}
\end{figure}

In the process of the calculations it is possible to visualize
intermediate results of modelling in real time, which allows to
control the progress of calculations. The modelling results can be
presented graphically. The user must tick the checkbox in the
table ``Modeling results'' and click next on ``Reports'' button in
the toolbar. Fig.~\ref{fig2} shows an example of the flow curves
for various FCC metals (aluminum, copper, nickel) at the
temperature $T=293K$.

Modelling results and the values of the model parameters used are
saved in a local database automatically. Information regarding the
progress of numerical solution of the task is also recorded in the
database. This information allows to generate recommendations for
further computer experiments.

The SPFCC software provides possibility to compare modelling
results for various applied effects in tables and graphical form.
According to modelling results, dependencies between any variables
as well as temperature and strain rate dependency can be
represented graphically.

\section{Illustrative example}

As an example of the model described above we show modelling
results for the case of constant strain rate and its comparison
with experimentally measured data. In case of the deformation of
crystal with a constant strain rate, the equation of applied
influence has the following form: $\dot{a}=const$. This is a
transcendental equation, which allows to find the stress at any
given time of the deformation.

\begin{figure}[!t]
\centering
\includegraphics[width=6in]{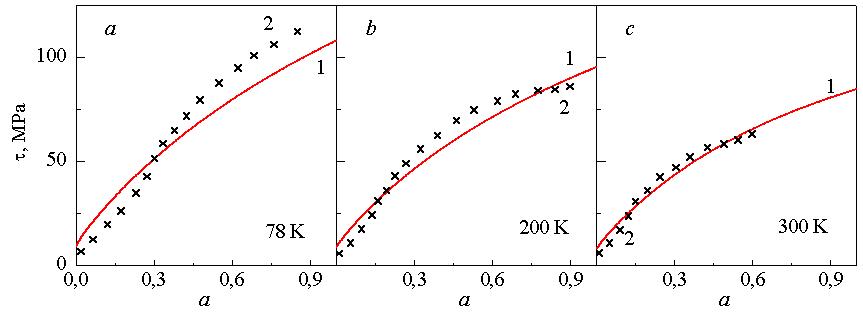}
\caption{Flow stress for copper monocrystal: 1 -- model data
(solid line); 2 -- experimental data (cross) \cite{Seeger1957}.
The numbers near curves are the deformation temperature given in
Kelvin degrees. The strain rate is $10^{-4} s^{-1}$.}\label{fig3}
\end{figure}

Using the above-mentioned equations of deformation defects balance
(\ref{eq17}) and the equation defining strain rate (\ref{eq18}),
calculations were performed for FCC metals with the values typical
for copper: $b=2.5 \cdot 10^{10} m$,  $\nu_D = 10^{13} s^{-1}$,
$\alpha= 0.5$, $\tau_f = 1 MPa$, $\nu=1/3$, $k = 1.38 \cdot
10^{-23} J/K$, $\omega_s=0.3$. 
The initial density of shear-forming dislocations were chosen to
be $10^{12}~m^{-2}$, and the initial density of dislocations in
vacancy and interstitial dipole configurations and the
concentration of interstitial atoms, vacancies, and divacancies
were chosen to be zero. Fig.~\ref{fig3} shows strain hardening
curves for copper monocrystal at various temperatures (solid
lines) in comparison with experimental data \cite{Seeger1957}. A
good agreement between experimental data and simulated curves is
obtained.

\begin{figure}[!h]
\centering
\includegraphics[width=6in]{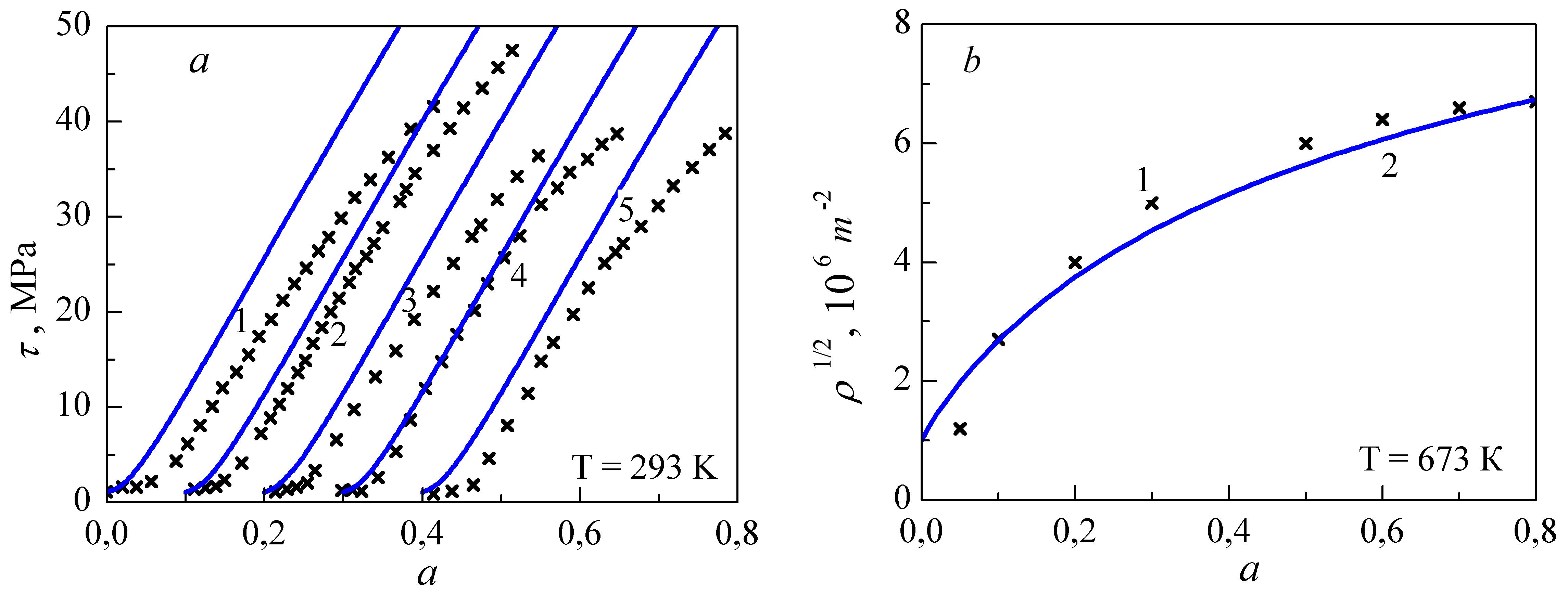}
\caption{Flow stress for copper monocrystal: model data (solid
line) and experimental data (cross) \cite{Basinski1964}. The
numbers near curves are the strain rate given in $s^{-1}$: 1 --
$1.27 \cdot 10^{-2}$, 2 -- $4.18 \cdot 10^{-3}$, 3 -- $4.34\cdot
10^{-4}$, 4 -- $4.09\cdot 10^{-5}$, 5 -- $4.46 \cdot 10^{-6}$;
(b). The square root of the dislocation density plotted versus the
shear strain for copper monocrystal: 1 -- experimental data
\cite{Popov1984}, 2 -- model data. The initial dislocation density
is $10^{12}m^{-2}$ and the strain rate is $10^{-4}
s^{-1}$.}\label{fig4}
\end{figure}

Fig.~\ref{fig4}\emph{a} shows strain hardening curves for copper
monocrystal at various strain rate (solid lines) in in comparison
with experimental data \cite{Basinski1964}.
Fig.~\ref{fig4}\emph{b} shows the square root of the dislocation
density plotted versus the shear strain for copper monocrystal
(solid line) in comparison with experimental data
\cite{Popov1984}.

\section{Conclusion} 

In this work a mathematical model of plastic deformation in FCC
metals was presented. Various features of plastic deformation by
slip under various loading conditions can be described within this
model. The model was implemented in-house software SPFCC (Slip
Plasticity of Face-Centered Cubic). It provides graphic user
interface and supports a study of plastic deformation of FCC
metals.

The ``Slip Plasticity of Face-Centered Cubic'' software creates a
comfortable and friendly environment for computational experiments
while modelling plastic deformation processes in both FCC-metals
and the dispersion-strengthened materials under various loading
conditions. The proposed modelling technic automates many user
actions, allowing to explore a variety of factors influencing
plastic deformation of FCC metals and dispersion-strengthened
materials. The results of the calculation  are stored in the
database of the software and are available for future
reference.The software also includes a number of tools for
collaboration within the scientific community. Tabular
representation of the results of the calculations is designed to
export data to third-party software for further post-processing.

It should be noted that only the flow stress curves of copper with
constant strain rate were considered in this work as an example.
It was verified that the proposed mathematical model describes the
plasticity of the FCC metals at various temperatures and strain
rates accurately, and the results of computer simulation are in a
good agreement with the experimental data. This computational
investigation provides a fundamental understanding of the
processes of intensity generation and annihilation of deformation
defects of the centered cubic metals for constant strain rate.
It's also demonstrates the influence of the typical parameters
values of FCC metals on the plastic deformation features.


\section{Acknowledgements}

This work was supported by the MSE Program ``Nauka'', contract
No.~1.604.2011. We thank Dr Andrey~Molotnikov (Monash University,
Australia) for critical reading and editorial suggestions for the
manuscript.



\appendix

\begin{table}[!ht]
\caption{Description of Mathematical Model Variables and
Parameters} \label{table1} \centering
\begin{tabular}{c m{4in}}
\hline \hline
Parameter & Description \\
\hline
$\rho$   & total dislocation density \\
$\rho_d^i$ & density of dislocation in the dipole configurations of the interstitial type  \\
$\rho_d^v$ & density of dislocation in the dipole configurations of the vacancy type \\
$\rho_d$ & $=\rho_d^v+\rho_d^i$, density of
dislocations in dipole configurations\\
$\rho_m$ & density of shear-forming dislocations \\
$c_{1v}$ & concentrations of monovacancies \\
$c_{2v}$ & concentrations of divacancies \\
$c_i$  & concentrations of interstitial atoms \\
$a$ & shear strain   \\
$\dot{a}$ & strain rate\\
$B$ & parameter, which is determined by the probability of dislocation barriers limiting the shear zone 
\\
$F$ & parameter, which is determined by the shape of dislocation loops and their distribution in the slip zone 
\\
$q$ & parameter that determines the intensity of point defects generation \\
$T$ & temperature \\
$\alpha$ & parameter of dislocations interaction \\
$b$ & Burgers vector \\
$D$ & diameter of the slip zone  \\
$G$ & shear modulus\\
$k$ & Boltzmann constant  \\
$\nu$ & Poisson coefficient  \\
$\omega_s$ & fraction of screw dislocations  \\
$P_{as}$ & probability of annihilation of screw dislocations \\
$Q_k$ & diffusion coefficient of point defects of the $k$th type $k =\{i, 1v, 2v\}$ \\
$r_a$ & critical capture radius \\
$\tau$ & flow stress  \\
$\tau_{dyn}$ & stress excess over the static resistance to dislocation motion \\
$\tau_f$ & friction stress \\
$\tau_a$ & athermic component of the resistance to the dislocation gliding  \\
\\
\hline \hline
\end{tabular}
\end{table}





\begin{thebibliography}{00}



\bibitem{Kuhlmann1962} D.~Kuhlmann-Wilsdorf, Transact. Metall. Soc. AIME. 224 (1962) 1047--1061.
\bibitem{Preston2003} D.L.~Preston, D.L.~Tonks, D.C.~Wallace,  J. of App. Phys. 93 (2003) 211--220.
\bibitem{Kocks03} U.~Kocks and H.~Mecking, Prog. in Mater. Sci. 48 (2003) 171--273.
\bibitem{Kolupaeva10} S.N.~Kolupaeva, T.A.~Kovalevskaya, O.I.~Daneyko, M.E.~Semenov, N.A.~Kulaeva, Bull. Russian Acad.
Sci.: Phys. 74 (2010) 1527--1531.
\bibitem{Zhan2011} H.F.~Zhan, Y.T.~Gu, C.~Yan, X.Q.~Feng, P.K.D.V.~Yarlagadda, Comput. Mater. Sci. 50 (2011) 3425--3430.
\bibitem{Lagneborg1972} R.~Lagneborg, Intern. Metals. Rev. 17 (1972) 130--146.
\bibitem{Essmann1979} V.~Essmann, H.~Mughrabi, Phil. Mag. (a) 40 (1979) 731--756.

\bibitem{Mecking1981} M.~Mecking, U.F.~Kocks, Acta Metal. 29 (1981) 1865--1875.
\bibitem{Estrin1998} Y.~Estrin, L.S.~Toth, A.~Molinari, Y.~Brechet, Acta Mater. 46 (1998) 5509--5522.
\bibitem{Popov2000} L.E.~Popov, S.N.~Kolupaeva, N.A.~Vihor, Comput. Mater. Sci. 19 (2000) 158--165.

\bibitem{Zerilli1987} F.J.~Zerilli, R.W.~Armstrong, J. of App. Phys. 5 (1987) 1816--1825.

\bibitem{Voyiadjis2008} G.Z.~Voyiadjis, A.H.~Almasri, Mechanics of Materials, 40 (2008) 549--563.

\bibitem{Chinh2010} N.Q.~Chinh, T.~Csana´di, J.~Gubicza, T.G.~Langdon, Acta Materialia 58 (2010) 5015–-5021.

\bibitem{Popov2000a} L.E.~Popov, S.N.~Kolupaeva, N.A.~Vihor, S.I.~Puspescheva, Comput. Mater. Sci. 19 (2000) 267--274.
\bibitem{Orlov1966} A.K.~Orlov, Kinetics of dislocations, in: Theory of crystals defects, Publishing House of
the Czechoslovak Acad. Sci., Prague, 1966, pp. 317--338.
\bibitem{Kolupaeva2005} S.N.~Kolupaeva, S.I.~Puspesheva, M.E.~Semenov, Proc. book of 11-th Inter. Conf. on Fracture. Turin (Italy) (2005)
847.
\bibitem{Semenov2010} M.E.~Semenov, S.N.~Kolupaeva, Tomsk Polytechnic University Bull. 317 (2010) 16--22.
\bibitem{Popov1982} L.E.~Popov, V.S.~Kobytev, T.A.~Kovalevskaya, Izv. Vyssh. Uchebn. Zaved. Fizika, 6 (1982) 56--82.
\bibitem{Indenbom1966} V.L.~Indenbom, Theory of dislocations -- present state and future, in: Theory of crystals
defects, Publishing House of the Czechoslovak Acad. Sci., Prague,
1966, pp. 2--16.
\bibitem{Gear1971} C.W.~Gear, Numerical Initial Value Problem in Ordinary Differential Equations. First ed., Prentice-Hall Inc.,
1971.
\bibitem{Seeger1957} A.~Seeger, J.~Diehl, S.~Mader, H.~Rebstock, Phil. mag. 15 (1957) 323--350.

\bibitem{Basinski1964} Z.~Basinski, S.~Basinski, Phil. Mag. (1964) 51-–80.

\bibitem{Popov1984} L.E.~Popov, V.S.~Kobytev, T.A.~Kovalevskaya, Plastic deformation of allows. Moscow, 1984.

\end{thebibliography}



\end{document}